\documentclass{myws-p10x7}

\begin{document}
\def\beq{\begin{equation}}
\def\eeq{\end{equation}}
\def\bea{\begin{eqnarray}}
\def\eea{\end{eqnarray}}
\def\bq{\begin{quote}}
\def\eq{\end{quote}}

\def\AJ{{\it Astrophys.J.} }
\def\AJL{{\it Ap.J.Lett.} }
\def\AJS{{\it Ap.J.Supp.} }
\def\AM{{\it Ann.Math.} }
\def\AP{{\it Ann.Phys.} }
\def\APJ{{\it Ap.J.} }
\def\APP{{\it Acta Phys.Pol.} }
\def\ASAS{{\it Astron. and Astrophys.} }
\def\BAMS{{\it Bull.Am.Math.Soc.} }
\def\CMJ{{\it Czech.Math.J.} }
\def\CMP{{\it Commun.Math.Phys.} }
\def\FP{{\it Fortschr.Physik} }
\def\HPA{{\it Helv.Phys.Acta} }
\def\IJMP{{\it Int.J.Mod.Phys.} }
\def\JMM{{\it J.Math.Mech.} }
\def\JP{{\it J.Phys.} }
\def\JCP{{\it J.Chem.Phys.} }
\def\LNC{{\it Lett. Nuovo Cimento} }
\def\SNC{{\it Suppl. Nuovo Cimento} }
\def\MPL{{\it Mod.Phys.Lett.} }
\def\NAT{{\it Nature} }
\def\NC{{\it Nuovo Cimento} }
\def\NP{{\it Nucl.Phys.} }
\def\PL{{\it Phys.Lett.} }
\def\PR{{\it Phys.Rev.} }
\def\PRL{{\it Phys.Rev.Lett.} }
\def\PRTS{{\it Physics Reports} }
\def\PS{{\it Physica Scripta} }
\def\PTP{{\it Progr.Theor.Phys.} }
\def\RMPA{{\it Rev.Math.Pure Appl.} }
\def\RNC{{\it Rivista del Nuovo Cimento} }
\def\SJPN{{\it Soviet J.Part.Nucl.} }
\def\SP{{\it Soviet.Phys.} }
\def\TMF{{\it Teor.Mat.Fiz.} }
\def\TMP{{\it Theor.Math.Phys.} }
\def\YF{{\it Yadernaya Fizika} }
\def\ZETF{{\it Zh.Eksp.Teor.Fiz.} }
\def\ZP{{\it Z.Phys.} }
\def\ZMP{{\it Z.Math.Phys.} }

\parskip 0.3cm

\def\gappeq{\mathrel{\rlap {\raise.5ex\hbox{$>$}}
{\lower.5ex\hbox{$\sim$}}}}

\def\lappeq{\mathrel{\rlap{\raise.5ex\hbox{$<$}}
{\lower.5ex\hbox{$\sim$}}}}

\def\Toprel#1\over#2{\mathrel{\mathop{#2}\limits^{#1}}}
\def\FF{\Toprel{\hbox{$\scriptscriptstyle(-)$}}\over{$\nu$}}

\title{Decline and Fall of the Standard Model?}

\author{John Ellis}

\address{Theoretical Physics Division, CERN,
        CH-1211 Geneva 23\\E-mail: John.Ellis@cern.ch}



\twocolumn[\maketitle\abstract{
Motivations for physics beyond the Standard Model are reviewed, with
particular emphasis on supersymmetry at the TeV scale. Constraints on the
minimal supersymmetric extension of the Standard Model with universal soft
supersymmetry-breaking terms (CMSSM) are discussed. These are also
combined with the supersymmetric interpretation of the anomalous magnetic
moment of the muon. The prospects for observing supersymmetry at
accelerators are reviewed using benchmark scenarios to focus the
discussion. Prospects for other experiments are discussed, including the
detection of cold dark matter, $\mu \to e \gamma$ and related processes,
as well as proton decay. \\
~~\\
\begin{center}
CERN-TH/2001-275~~~~~~hep-ph/0110192
\end{center}
}]
\section{Introduction}

The empire of the Standard Model has resisted all attacks by accelerator
data. Nevertheless, we theorists are driven to overcome our ignorance of
the barbarian territory beyond its frontiers. In the gauge sector, the
Standard Model has three independent gauge couplings and (potentially) a
CP-violating phase in QCD. In the Yukawa sector, it has six random-seeming
quark masses, three charged-lepton masses, three weak mixing angles and
the Kobayashi-Maskawa phase. Finally, the symmetry-breaking sector has at
least two free parameters. Moreover, this list of 19 parameters in the
Standard Model begs the more fundamental questions of the origins of the
particle quantum numbers. As if this were not enough, non-accelerator
neutrino experiments~\cite{nus} now convince us that we need three
neutrino mass parameters, three neutrino mixing angles and three
CP-violating phases in the neutrino sector: one observable in oscillation
experiments and two that affect $\beta\beta_{0\nu}$ experiments, without
even talking about the mechanism of neutrino mass generation. Moreover, we
should not forget about gravity, with at least two parameters to
understand: Newton's constant $G_N \equiv m_P^{-2} \sim (10^{19}$
GeV)$^{-2}$ and the cosmological `constant', which recent data suggest is
non-zero~\cite{lambda}, and may not even be constant. Talking of 
cosmology, we would need at least one
extra parameter to produce an inflationary potential, and at least one
other to generate the baryon asymmetry, which cannot be explained within
the Standard Model.

Confronted by our ignorance of so much barbarian territory, we legions
of theorists organize our explorations on three main fronts: {\it
unification} -- the quest for a single framework for all gauge
interactions, {\it flavour} -- the quest for explanations of the
proliferation of quark and lepton types, their mixings and CP violating
phases, and {\it mass} -- the quest for the origin of particle masses and
an explanation why they are so much smaller than the Planck mass $m_P \sim
10^{19}$ GeV. Beyond all these beyonds, other scouting parties of
theorists seek a {\it Theory of Everything} that includes gravity,
reconciles it with quantum mechanics, explains the origin of space-time
and why we live in four dimensions (if we do so).

Physics beyond the Standard Model is therefore a very broad subject.
However, many aspects are discussed here by other speakers: electroweak
flavour physics~\cite{RB}, CP violation~\cite{MN}, the Higgs
sector~\cite{FZ}, $g_\mu - 2$~\cite{g-2}, searches for new
particles~\cite{GH}, neutrinos~\cite{HM}, dark matter~\cite{KKG}, strings
and extra dimensions~\cite{LR}. Therefore, in this talk I seek a
complementary approach.

For reasons that I describe in Section 2, many theorists believe that
supersymmetry is the inescapable framework for discussing physics at the
TeV scale and beyond. In the rest of this talk, I first discuss the
constraints imposed on (the simplest) supersymmetric models by the
available experimental and cosmological constraints, then address the
prospects for understanding $g_\mu - 2$ in supersymmetric models, the
prospects for detecting sparticles directly at present and future
colliders, and the prospects for non-collider experiments, including the
searches for dark matter, $\mu\rightarrow e\gamma$ and proton decay.

\section{The Electroweak Vacuum}

\begin{figure}
\epsfxsize180pt
\figurebox{120pt}{160pt}{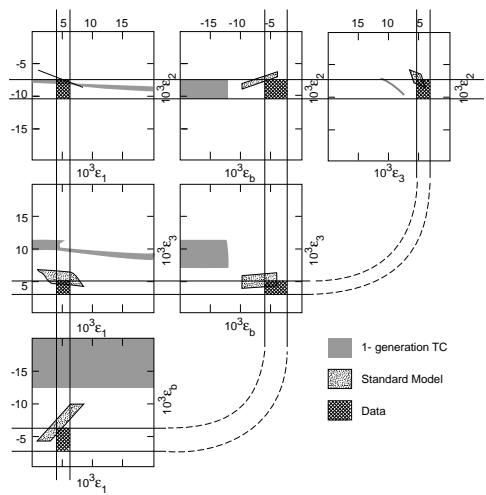} 
\caption[]{Predictions for the radiative corrections $\epsilon_i$
in the Standard Model and a minimal one-generation model~\cite{EFL} are 
compared with the precision electroweak data~\cite{Aetal}. 
\label{fig:TC}}
\end{figure}

The generation of particle masses requires the breaking of gauge symmetry in
the vacuum:
\beq
m_{W,Z} \not= 0 \Leftrightarrow <0\vert X_{I,I_3}\vert 0> \not= 0
\label{one}
\eeq
for some field $X$ with isospin $I$ and third component $I_3$. The measured
ratio
\beq
\rho \equiv {m^2_W\over m^2_Z ~\cos^2\theta_W} \simeq 1
\label{two}
\eeq
tells us that $X$ mainly has $I = 1/2$~\cite{RV}, which is also what is 
needed to
generate fermion masses. The key question is the nature of the field $X$:
is it elementary or composite? A fermion-antifermion condensate $v \equiv
<0\vert X\vert 0> = <0\vert\bar FF\vert 0> \not= 0$ would be analogous to
what we know from QCD, where $<0\vert\bar qq\vert 0> \not= 0$, and
conventional superconductivity, where $<0\vert e^-e^-\vert 0> \not= 0$.
However, analogous `technicolour' models of electroweak symmetry 
breaking~\cite{TC}
fail to fit the values of the radiative corrections $\epsilon_i$ to $\rho$
and other quantities extracted from the precision electroweak data
provided by LEP and other experiments, as seen in 
Fig.~\ref{fig:TC}~\cite{Aetal}. One 
cannot
exclude the possibility that some calculable variant of technicolour might
emerge that is consistent with the data, but for now we focus on
elementary Higgs models.

\begin{figure}
\epsfxsize180pt
\figurebox{120pt}{160pt}{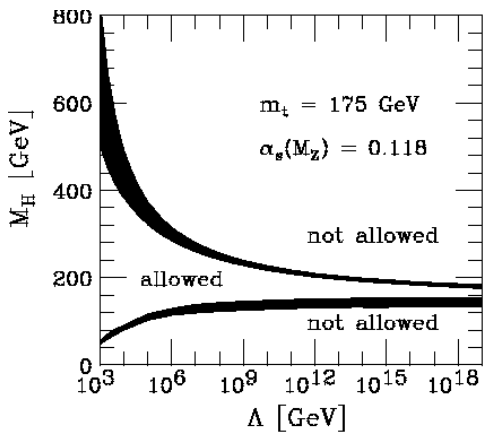}
\caption[]{The range allowed for the mass of the Higgs boson if the 
Standard Model is to remain valid up to a given scale $\Lambda$. In the 
upper part of the plane, the effective potential blows up, whereas in the 
lower part the present electroweak vacuum is unstable~\cite{tdown}.} 
\label{fig:HR} 
\end{figure}

Within this framework, the data favour a relatively light Higgs boson,
with $m_H \simeq$ 115 GeV, just above the exclusion unit
provided by direct searches at LEP, being the
`most-probable'~\cite{precision}. This is one reason why many theorists
were excited by the possible sighting during the last days of LEP of a
Higgs boson, with a preferred mass of 115.6 GeV~\cite{GH}. If this were to
be confirmed, it would suggest that the Standard Model breaks down at some
relatively low energy $\lappeq 10^3$ TeV~\cite{ER}. As seen in
Fig.~\ref{fig:HR}, above this scale the effective Higgs potential of the
Standard Model becomes unstable as the quartic Higgs self-coupling is
driven negative by radiative corrections due to the relatively heavy top
quark~\cite{tdown}. This is not necessarily a disaster, and it is possible
that the present electroweak vacuum might be metastable, provided that its
lifetime is longer than the age of the Universe~\cite{Setal}. However, we
would surely feel more secure if such instability could be avoided.

\begin{figure}
\epsfxsize180pt
\figurebox{120pt}{160pt}{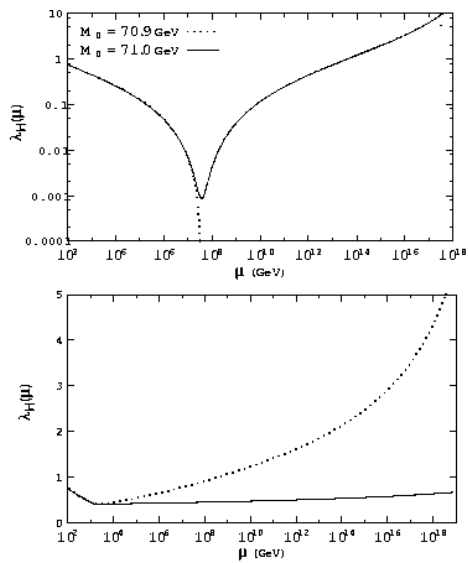}   
\epsfxsize180pt
\caption[]{(a) If the quartic coupling $M_0$ (\ref{three}) is too large, the 
effective potential blows up (solid line), whereas it is unstable if $M_0$
is too 
small (dotted line), indicating a need for fine tuning. (b) This occurs
naturally in a 
supersymmetric model (solid line) but not if the $\tilde H$ are omitted 
(dotted line)~\cite{ER}.} 
\label{fig:ER} 
\end{figure}

This may be done by introducing new bosons $\phi$ coupled to the Higgs
field~\cite{ER}:
\beq
\lambda_{22}\vert H\vert^2~\vert\phi\vert^2~~:~~M^2_0 \equiv \lambda_{22}
v^2
\label{three}
\eeq
As seen in Fig.~\ref{fig:ER}a, the effective potential is very sensitive 
to the
coupling parameter $M_0$: for $M_0 \leq$ 70.9 GeV in this example, the
potential still collapses, whereas for $M_0 \geq$ 71.0 GeV the potential
blows up instead. Thus the bosonic coupling (\ref{three}) must be finely
tuned~\cite{ER}. This occurs naturally in supersymmetry, in which the 
Higgs bosons
are accompanied by fermionic partners $\tilde H$. As seen in 
Fig.~\ref{fig:ER}b,
again the Higgs coupling blows up in the absence of the $\tilde H$,
whereas it is well behaved in the minimal supersymmetric extension of the
Standard Model (MSSM).

The avoidance of fine tuning has long been the primary motivation for
supersymmetry at the TeV scale~\cite{hierarchy}. This issue is normally 
formulated in
connection with the hierarchy problem: why/how is $m_W \ll m_P$, or
equivalently why is $G_F \sim 1/m^2_W \gg G_N = 1/m^2_P$, or equivalently
why does the Coulomb potential in an atom dominate over the Newton
potential, $e^2 \gg G_N m_p m_e \sim (m/m_P)^2$, where $m_{p,e}$ are the
proton and electron masses? One might think naively that it would be
sufficient to set $m_W \ll m_P$ by hand. However, radiative corrections
tend to destroy this hierarchy. For example, one-loop diagrams generate
\beq
\delta m^2_W = {\cal O}\left({\alpha\over\pi}\right)~\Lambda^2 \gg m^2_W
\label{four}
\eeq
where $\Lambda$ is a cut-off representing the appearance of new physics, and
the
inequality in (\ref{four}) applies if $\Lambda\sim 10^3$ TeV, and even 
more so if  $\Lambda \sim m_{GUT} \sim
10^{16}$
GeV or $ \sim m_P \sim 10^{19}$ GeV. If the radiative corrections to a 
physical
quantity
are much larger than its measured values, obtaining the latter requires
strong
cancellations, which in general require fine tuning of the bare input
parameters.
However, the necessary cancellations are natural in supersymmetry, where 
one has
equal
numbers of bosons $B$ and fermions $F$ with equal couplings, so that
(\ref{four})
is replaced by
\beq
\delta m^2_W = {\cal O}\left({\alpha\over\pi}\right)~\vert m^2_B - 
m^2_F\vert~.
\label{five}
\eeq
The residual radiative correction is naturally small if
\beq
\vert m^2_B - m^2_F\vert \lappeq 1~{\rm TeV}^2
\label{six}
\eeq
Note that this argument is logically distinct from that in the previous
paragraph. There supersymmetry was motivated by the control of logarithmic
divergences, and here by the absence of quadratic divergences.

\section{The MSSM}

The MSSM has the same gauge interactions as the Standard Model, and
similar Yukawa couplings. A key difference is the necessity of two Higgs
doublets, in order to give masses to all the quarks and leptons, and to
cancel triangle anomalies. This duplication is important for
phenomenology: it means that there are five physical Higgs bosons, two
charged $H^\pm$ and three neutral $h, H, A$. Their quartic
self-interactions are determined by the gauge interactions, solving the
vacuum instability problem mentioned above and limiting the possible mass
of the lightest neutral Higgs boson. However, the doubling of the Higgs
multiplets introduces two new parameters: $\tan\beta$, the ratio of Higgs
vacuum expectation values and $\mu$, a parameter mixing the two Higgs
doublets.

There are two key experimental hints in favour of supersymmetry. One is
provided by the LEP measurements of the gauge couplings, that are in very
good agreement with supersymmetric GUTs~\cite{susyGUTs} if sparticles 
weigh $\sim 1$~TeV. This agreement
appears completely fortuitous in composite Higgs models~\cite{TC}, and is 
difficult
(though not impossible~\cite{DDG}) to reproduce accurately in models with 
large extra
dimensions~\cite{GR}. The other experimental hint is provided by the
preference of the precision electroweak data for a relatively light Higgs
boson~\cite{precision}. In the MSSM, one predicts $m_h \lappeq$ 130
GeV~\cite{Hmass,FZ}, right in the preferred range, whereas composite Higgs
model generally predict heavier effective Higgs masses.

The gauge symmetries of the MSSM would permit the inclusion of interactions
that violate baryon number and/or lepton number~\cite{Rviol}:
\beq
\lambda LLE^c + \lambda^\prime QD^cL + \lambda^" U^cD^cD^c
\label{seven}
\eeq
where the $L(Q)$ are left-handed lepton (quark) doublets and the
$E^c(D^c,U^d)$ are conjugates of the right-handed lepton (quark) singlets.
Their possible appearance is ignored in this talk, in which case the
lightest supersymmetric particle is stable, and hence a candidate for dark
matter~\cite{EHNOS}. In the following this is assumed to be a neutralino, 
i.e., a
mixture of the $\tilde\gamma, \tilde H$ and $\tilde Z$.

The final ingredient in the MSSM is the soft supersymmetry breaking, in
the form of scalar masses $m_0$, gaugino masses $m_{1/2}$ and trilinear
couplings $A$~\cite{DG}. These are presumed to be inputs from physics at 
some
high-energy scale, e.g., from some supergravity or superstring theory,
which then evolve down to lower energy scale according to well-known
renormalization-group equations. In the case of the Higgs multiplets, this
renormalization can drive the effective mass-squared negative, triggering
electroweak symmetry weaking~\cite{renn}. In this talk, it is assumed that 
the $m_0$
are universal at the input scale~\footnote{Universality between the
squarks and sleptons of different generations is motivated by upper limits
on flavour-changing neutral interactions~\cite{FCNC}, but universality 
between the
soft masses of the $L, E^c, Q^c, D^c$ and $U^c$ is not so well
motivated.}, as are the $m_{1/2}$ and $A$ parameters. In this case the
free parameters are
\beq
m_0, m_{1/2}, A \quad {\rm and}\quad \tan\beta~,
\label{eight}
\eeq
with $\mu$ being determined by the electroweak vacuum conditions, up to a 
sign.

This constrained MSSM (CMSSM) serves \\
as the basis for the subsequent
discussion. It has the merit of being sufficiently specific that the
different phenomenological constraints can be combined meaningfully. On
the other hand, it is just one of the phenomenological possibilities
offered by supersymmetry~\cite{othersusy}.

\section{Constraints on the CMSSM}

\begin{figure}
\epsfxsize200pt
\figurebox{120pt}{160pt}{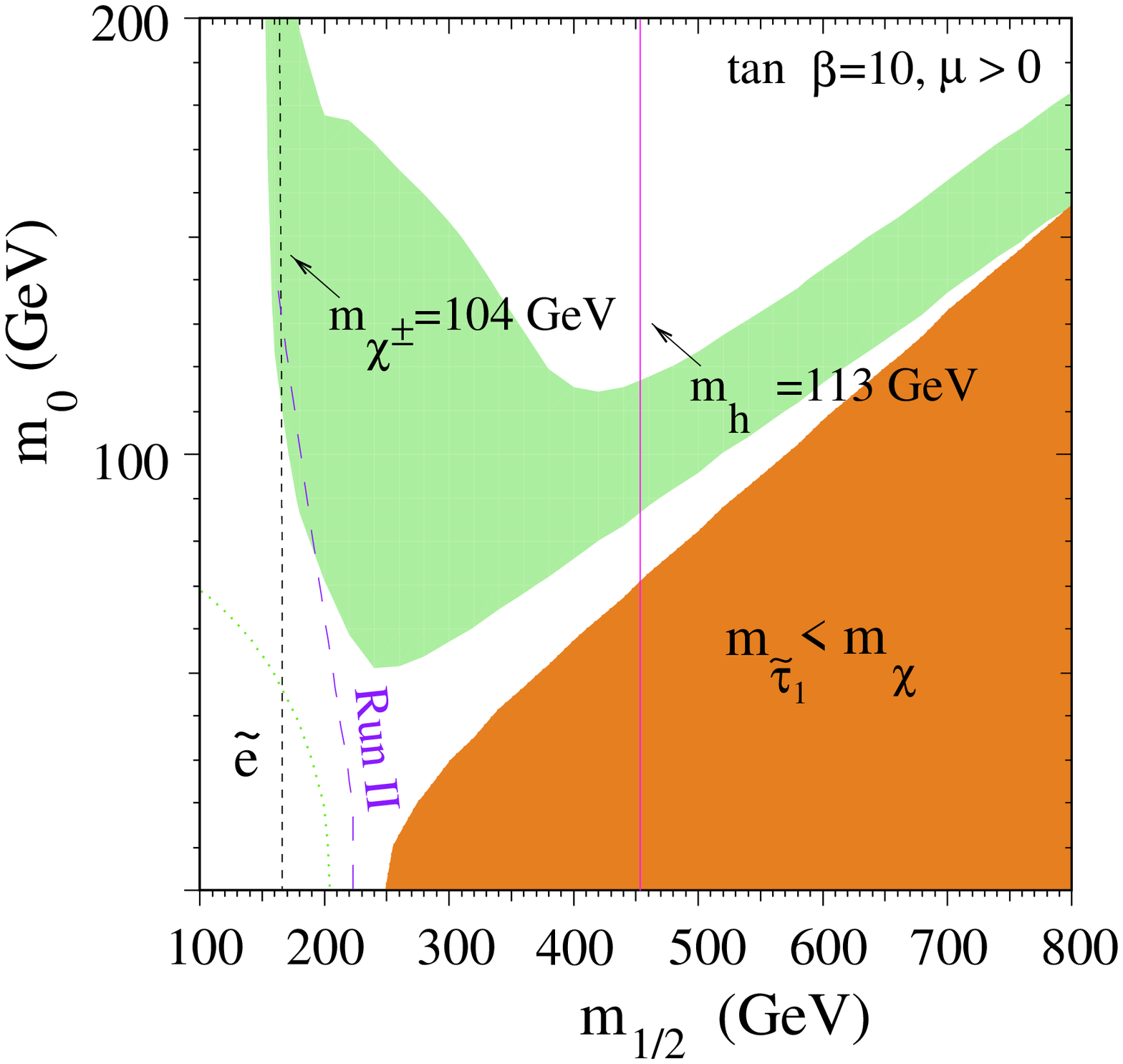}
\epsfxsize200pt
\figurebox{120pt}{160pt}{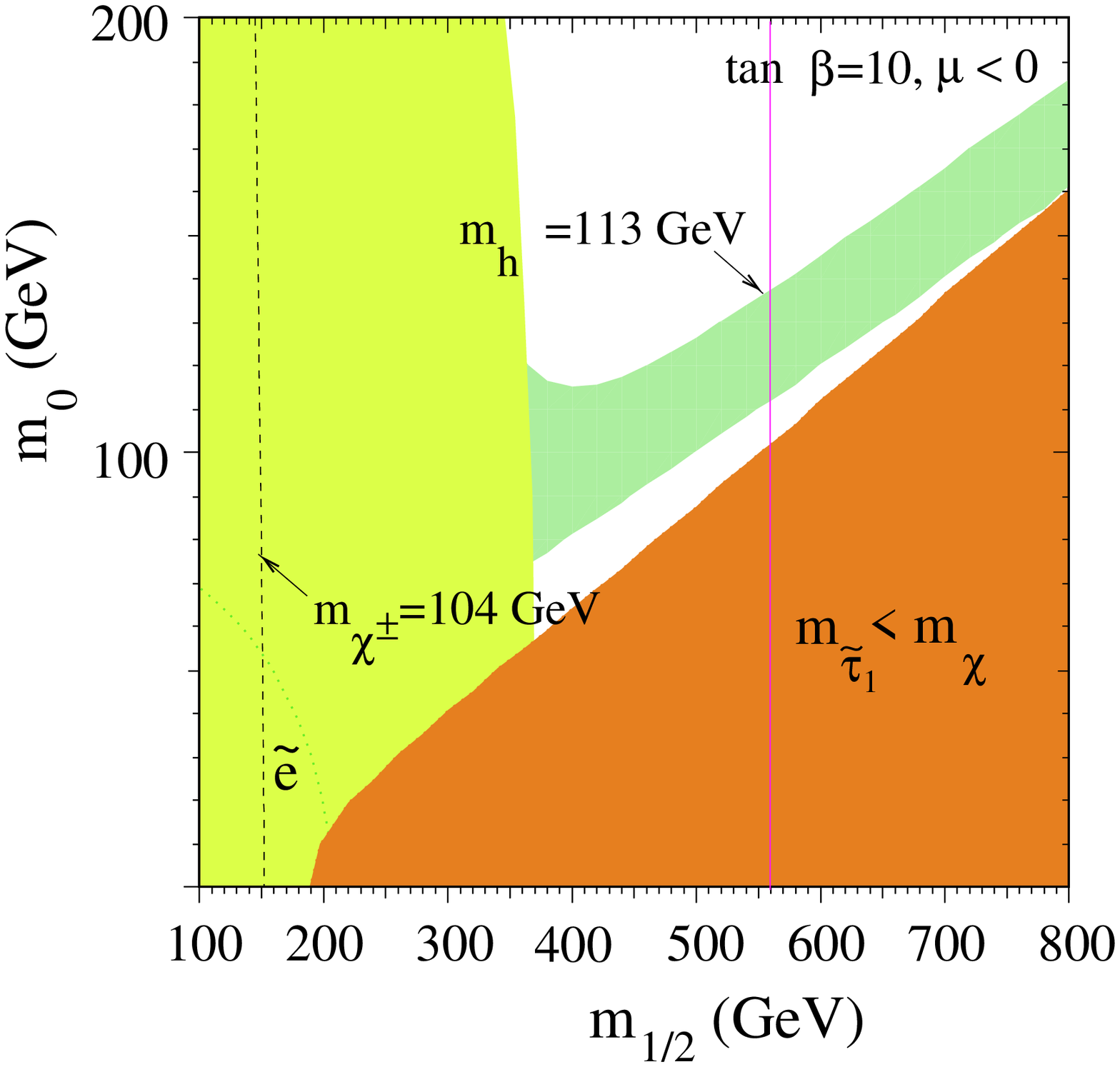}
\caption[]{Compilations of phenomenological constraints on the CMSSM for 
$\tan \beta = 10$ and (a) $\mu > 0$, (b) $\mu < 0$. Representative 
contours of the selectron, chargino and Higgs masses are indicated, as 
is the likely physics reach of Run II of the Tevatron Collider in (a). The
dark 
shaded regions are excluded because the LSP is charged, whereas a 
neutralino LSP has acceptable relic density (\ref{ten}) in the 
light-shaded regions~\cite{EGNO}. The medium-shaded region in (b) is
excluded by $b \to s \gamma$~\cite{bsg}.} 
\label{fig:CMSSM} 
\end{figure}

Important constraints on the CMSSM parameter space are provided by direct
searches at LEP and the Tevatron collider~\cite{GH}, as seen in
Fig.~\ref{fig:CMSSM}. One of these is the limit $m_{\chi^\pm} \gappeq$ 103
GeV provided by chargino searches at LEP, where the third significant
figure depends on other CMSSM parameters. LEP has also provided lower
limits on slepton masses, of which the strongest is $m_{\tilde e}\gappeq$
99 GeV, again depending only sightly on the other CMSSM parameters, as
long as $m_{\tilde e} - m_\chi \gappeq$ 10 GeV. The most important
constraints on the $u, d, s, c, b$ squarks and gluinos are provided by the
Tevatron collider: for equal masses $m_{\tilde q} = m_{\tilde g} \gappeq$
300 GeV. In the case of the $\tilde t$, LEP provides the most stringent
limit when $m_{\tilde t} - m_\chi$ is small, and the Tevatron for larger
$m_{\tilde t} - m_\chi$. Their effect is almost to exclude the range of
parameter space where electroweak baryogenesis is possible~\cite{leptog}.

Another important constraint is provided by the LEP limit on the Higgs mass:
$m_H
> $ 114.1 GeV. This holds in the Standard Model, for the lightest Higgs 
boson $h$
in the general MSSM for $\tan\beta \lappeq 5$, and in the CMSSM for all
$\tan\beta$, at least as long as CP is conserved~\footnote{The lower 
bound on the lightest MSSM Higgs boson may be relaxed significantly if 
CP violation feeds into the MSSM Higgs sector~\cite{CEPW}.}. Since $m_h$ 
is sensitive 
to sparticle masses, particularly $m_{\tilde t}$, via loop corrections:
\beq
\delta m^2_h \propto {m^4_t\over m^2_W}~\ln\left({m^2_{\tilde t}\over
m^2_t}\right)~ + \ldots
\label{nine}
\eeq
the Higgs limit also imposes important constraints on the CMSSM parameters,
principally $m_{1/2}$ as seen in Fig.~\ref{fig:CMSSM}.

Also shown in Fig.~\ref{fig:CMSSM} is the constraint imposed by
measurements of $b\rightarrow s\gamma$~\cite{bsg}. These agree with the
Standard Model, and therefore provide bounds on chargino and charged Higgs
masses, for example. For moderate $\tan\beta$, the $b\rightarrow s\gamma$
constraint is more important for $\mu < 0$, as seen in
Fig.~\ref{fig:CMSSM}b, but it is also significant for $\mu > 0$ when
$\tan\beta$ is large.

Fig.~\ref{fig:CMSSM} also displays the regions where the supersymmetric 
relic density $\rho_\chi = \Omega_\chi \rho_{critical}$ falls within the 
preferred range
\beq
0.1 < \Omega_\chi h^2 < 0.3
\label{ten}
\eeq
The upper limit is rigorous, since astrophysics and cosmology tell us that
the
total matter density $\Omega_m \lappeq 0.4$, and the Hubble expansion rate
$h
\sim 1/\sqrt{2}$ to within about 10 \% (in units of km/s/Mpc). On the 
other hand, the lower limit
in
(\ref{ten}) is optional, since there could be other important contributions
to
the overall matter density.

\begin{figure}
\epsfxsize200pt
\figurebox{120pt}{160pt}{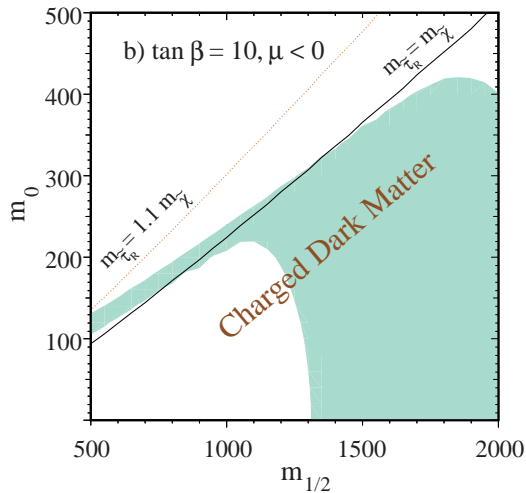}
\caption[]{The large-$m_{1/2}$ `tail' of the $\chi - {\tilde \tau_1}$ 
coannihilation region 
for $\tan \beta = 10$ and $\mu < 0$~\cite{ourcoann}.}
\label{fig:coann}
\end{figure}

\begin{figure}
\epsfxsize200pt 
\figurebox{120pt}{160pt}{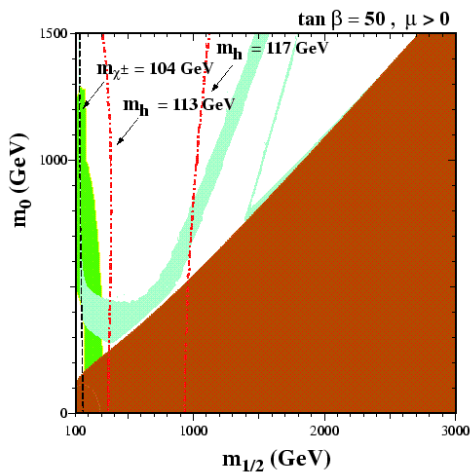}
\caption[]{The region where the cosmological relic density is in the 
preferred range (\ref{ten}) for $\tan \beta = 50$ and $\mu > 0$. Note 
the rapid-annihilation `funnel' at intermediate $m_0 / 
m_{1/2}$~\cite{EFGOSi}.} 
\label{fig:funnel} 
\end{figure}

\begin{figure}
\epsfxsize180pt 
\figurebox{120pt}{160pt}{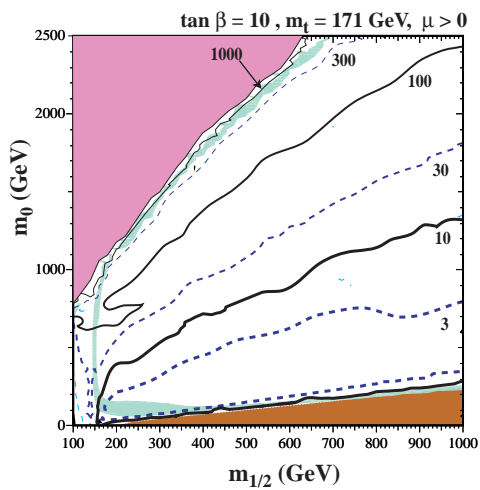}
\caption[]{The $m_{1/2}, m_0$ plane for $\tan \beta = 10$ and $\mu > 0$, 
including the `focus-point' region~\cite{focus} at large $m_0$, close to
the boundary of the shaded region where electroweak symmetry breaking
occurs, and exhibiting 
contours of the cosmological sensitivity (\ref{twelve})~\cite{EO}.} 
\label{fig:EO}
\end{figure}

As is seen in Fig.~\ref{fig:CMSSM}, there are generic regions of the CMSSM 
parameter space
where the relic density falls within the preferred range (\ref{ten}). What
goes
into the calculation of the relic density? It is controlled by the
annihilation
rate~\cite{EHNOS}:
\beq
\rho_\chi = m_\chi n_\chi : n_\chi \sim {1\over
\sigma_{ann}(\chi\chi\rightarrow\ldots)}
\label{eleven}
\eeq
and the typical annihilation rate $\sigma_{ann} \sim 1/m_\chi^2$. For this
reason, the relic density typically increases with the relic mass, and
this combined with the upper bound in (\ref{ten}) then leads to the common
expectation that $m_\chi \lappeq$ 1 TeV. However, there are various ways
in which the generic upper bound on $m_\chi$ can be increased along
filaments in the $(m_{1/2},m_0)$ plane. For example, if the
next-to-lightest sparticle (NLSP) is not much heavier than $\chi$: $\Delta
m/m_\chi \lappeq 0.1$, the relic density may be suppressed by
coannihilation: $\sigma (\chi + $NLSP$ \rightarrow \ldots )$~\cite{coann}. 
In this way,
the allowed CMSSM region may acquire a `tail' extending to large $m_\chi$,
as in the case where the NLSP is the lighter stau: $\tilde\tau_1$ and
$m_{\tilde\tau_1} \sim m_\chi$ as seen in 
Fig.~\ref{fig:coann}~\cite{ourcoann}. Another mechanism for
extending the allowed CMSSM region to large $m_\chi$ is rapid annihilation
via a direct-channel pole when $m_\chi \sim {1\over 2} m_{Higgs, 
Z}$~\cite{funnel,EFGOSi}. This
may yield a `funnel' extending to large $m_{1/2}$ and $m_0$ at large
$\tan\beta$, as seen in Fig.~\ref{fig:funnel}~\cite{EFGOSi}. Another 
allowed region at 
large $m_{1/2}$
and $m_0$ is the `focus-point' region~\cite{focus}, which is adjacent to 
the boundary
of the region where electroweak symmetry breaking is possible, as seen in
Fig.~\ref{fig:EO}. However, in this region $m_\chi$ is not particularly 
large.

These filaments extending the preferred CMSSM parameter space are clearly
exceptional, in some sense, so it is important to understand the sensitivity
of the relic density to input parameters, unknown higher-order effects, 
etc. One proposal is the relic-density fine-tuning measure~\cite{EO}
\beq
\Delta^\Omega \equiv \sqrt{\sum_i ~~{\partial\ln (\Omega_\chi h^2)\over
\partial
\ln a_i}}
\label{twelve}
\eeq
where the sum runs over the input parameters, which might include
(relatively) poorly-known Standard Model quantities such as $m_t$ and
$m_b$, as well as the CMSSM parameters $m_0, m_{1/2}$, etc. As seen in
Fig.~\ref{fig:EO}, the sensitivity $\Delta^\Omega$ (\ref{twelve}) is 
relatively small
in the `bulk' region at low $m_{1/2}$, $m_0$, and $\tan\beta$. However, it
is somewhat higher in the $\chi - \tilde\tau_1$ coannihilation `tail', and
at large $\tan\beta$ in general. The sensitivity measure $\Delta^\Omega$
(\ref{twelve}) is particularly high in the rapid-annihilation `funnel' and
in the `focus-point' region. This explains why published relic-density
calculations may differ in these regions~\cite{otherOmega}, whereas they 
agree well when
$\Delta^\Omega$ is small: differences may arise because of small
differences in the treatments of the inputs.

It is important to note that the relic-density fine-tuning measure
(\ref{twelve}) is distinct from the traditional measure of the fine-tuning
of the electroweak scale~\cite{EENZ}:
\beq
\Delta_i \equiv {\partial \ln m_W\over \partial \ln a_i}
\label{thirteen}
\eeq
This electroweak fine-tuning is a completely different issue, and values
of the $\Delta_i$ are not necessarily related to values of
$\Delta^\Omega$.  Electroweak fine-tuning is sometimes used as a criterion
for restricting the CMSSM parameters. However, the interpretation of the
$\Delta_i$ (\ref{thirteen}) is unclear. How large a value of $\Delta_i$ is
tolerable? Different physicists may well have different pain thresholds.
Moreover, correlations between input parameters may reduce its value in
specific models.

\section{Muon Anomalous Magnetic Moment}

As reported at this meeting~\cite{g-2}, the BNL E821 experiment has 
recently reported a
2.6-$\sigma$ deviation of $a_\mu\equiv {1\over 2} (g_\mu -2)$ from the
Standard Model prediction~\cite{BNL}:
\beq
a_\mu^{exp} - a_\mu^{th} = (43 \pm 16) \times 10^{-10}
\label{forteen}
\eeq
The largest contribution to the error in (\ref{forteen}) is the
statistical error of the experiment, which will soon be significantly
reduced, as many more data have already been recorded. The next-largest
error is that due to strong-interaction uncertainties in the Standard
Model prediction. Recent estimates converge on an estimate of about
7$\times 10^{-10}$ for the error in the hadronic vacuum polarization
constribution to (\ref{forteen})~\cite{DHSNTY}, and the error in the 
hadron
light-by-light scattering contribution is generally thought to be 
smaller~\cite{JB}.
Therefore, if the central value in (\ref{forteen}) does not change
substantially with the new data, this would be strong evidence for new
physics at the TeV scale.

\begin{figure}
\epsfxsize180pt
\figurebox{120pt}{160pt}{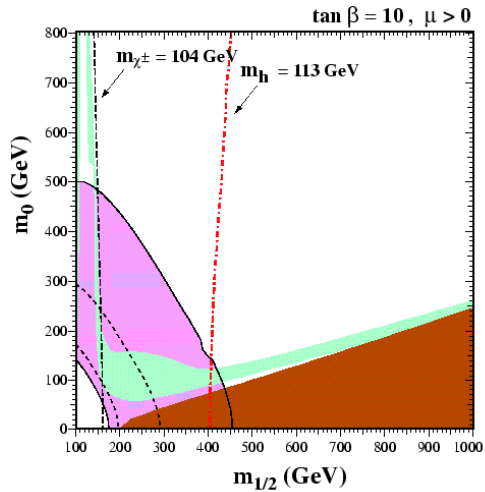}
\caption[]{The medium-shaded region is that compatible with the BNL E821 
measurement of $g_\mu - 2$ at the 2-$\sigma$ level~\cite{g-2,BNL}, the
light-shaded region has a relic density in the preferred range
(\ref{ten}), and the dark-shaded region does not have a neutralino LSP.
Good 
compatibility is found between $g_\mu - 2$ and the other phenomenological
constraints for $\tan \beta \sim 5$ or more~\cite{susygmu}.} 
\label{fig:g-2}
\end{figure}

As many authors have 
pointed 
out~\cite{susygmu}, the 
discrepancy 
(\ref{forteen}) could
well be explained by supersymmetry if $\mu > 0$ and $\tan\beta$ is not too
small, as exemplified in Fig.~\ref{fig:g-2}. Good consistency with all 
the experimental and cosmological
constraints on the CMSSM is found for $\tan\beta\lappeq 10$ and $m_\chi
\simeq$ 150 to 350 GeV. Already before the measurement (\ref{forteen}),
the LHC was thought to have a good chance of discovering 
supersymmetry~\cite{susyLHC}. If
the result (\ref{forteen}) were to be confirmed, this would be almost
guaranteed, as we now discuss.

\section{Prospects for Observing Supersymmetry at Accelerators}

\begin{figure}
\epsfxsize180pt
\figurebox{120pt}{160pt}{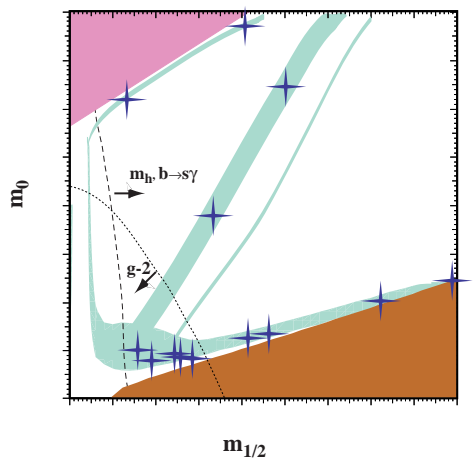}
\caption[]{
Schematic overview of the benchmark points proposed in~\cite{benchmark}.
They were chosen to be compatible with the indicated experimental
constraints, as well as have a relic density in the preferred range
(\ref{ten}). The points are intended to illustrate the range of available
possibilities.} 
\label{fig:Bench} 
\end{figure}

As an aid to the assessment of the prospects for detecting sparticles at
different accelerators, benchmark sets of supersymmetric parameters have
often been found useful~\cite{benchmarks}, since they provide a focus for
concentrated discussion. A set of post-LEP benchmark scenarios in the
CMSSM has recently been proposed~\cite{benchmark}, and are illustrated
schematically in Fig.~\ref{fig:Bench}. They take into account
the direct searches for sparticles and Higgs bosons, $b\rightarrow
s\gamma$ and the preferred cosmological density range (\ref{ten}). About a
half of the proposed benchmark points are consistent with $g_\mu -2$
(\ref{forteen}) at the $2-\sigma$ level, but this was not imposed as an
absolute requirement.

The proposed points were chosen not to provide an `unbiased' statistical
sampling of the CMSSM parameter space, whatever that means in the absence
of a plausible {\it a priori} measure, but rather are intended to
illustrate the different possibilities that are still allowed by the
present constraints~\cite{benchmark}. Five of the chosen points are in the
`bulk' region at small $m_{1/2}$ and $m_0$, four are spread along the
coannihilation `tail' at larger $m_{1/2}$ for various values of
$\tan\beta$, two are in the `focus-point' region at large $m_0$, and two
are in rapid-annihilation `funnels' at large $m_{1/2}$ and $m_0$. The
proposed points range over the allowed values of $\tan\beta$ between 5 and
50.  Most of them have $\mu > 0$, as favoured by $g_\mu - 2$, but there
are two points with $\mu < 0$.

Various derived quantities in these
supersymmetric benchmark scenarios, including the relic density, $g_\mu -
2, b \rightarrow s\gamma$, electroweak fine-tuning $\Delta$ and the
relic-density sensitivity $\Delta^\Omega$, are given in~\cite{benchmark}.
These enable
the reader to see at a glance which models would be excluded by which
refinement of the experimental value of $g_\mu - 2$. Likewise, if you find
some amount of fine-tuning uncomfortably large, then you are free to
discard the corresponding models.

\begin{figure}
\epsfxsize200pt 
\figurebox{120pt}{160pt}{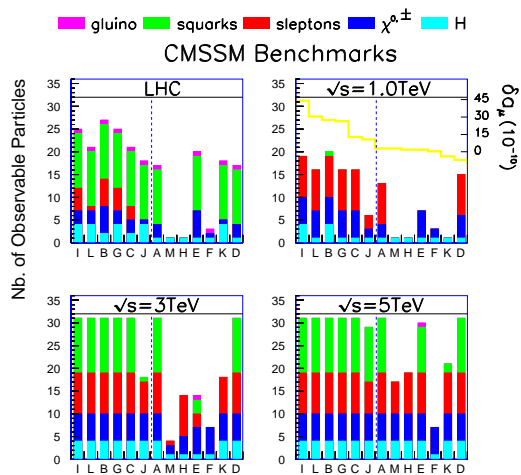}
\caption[]{Estimates of the numbers of different types of CMSSM particles 
that may be detectable~\cite{benchmark} at (a) the LHC, (b) a 1-TeV linear 
$e^+ e^-$ collider~\cite{LC}, and (c,d) a 3(5)-TeV $e^+ e^-$~\cite{CLIC}
or $\mu^+ 
\mu^-$ collider~\cite{MC}. Note the complementarity between the sparticles 
detectable at the LHC and at a 1-TeV linear $e^+ e^-$ collider.} 
\label{fig:Manhattan} 
\end{figure}

The LHC collaborations have analyzed their reach for sparticle detection
in both generic studies and specific benchmark scenarios proposed
previously~\cite{susyLHC}. Based on these studies,
Fig.~\ref{fig:Manhattan} displays estimates how many different sparticles
may be seen at the LHC in each of the newly-proposed benchmark
scenarios~\cite{benchmark}. The lightest Higgs boson is always found, and
squarks and gluinos are usually found, though there are some scenarios
where no sparticles are found at the LHC. The LHC often misses heavier
weakly-interacting sparticles such as charginos, neutralinos, sleptons and
the other Higgs bosons.

\begin{figure}
\epsfxsize200pt 
\figurebox{120pt}{160pt}{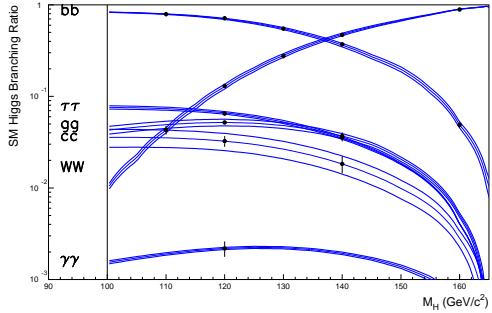}
\caption[]{Analysis of the accuracy with which Higgs decay branching 
ratios may be measured with a linear $e^+ e^-$ collider~\cite{BD}.}
\label{fig:Battaglia}
\end{figure}

The physics capabilities of linear $e^+e^-$ colliders are amply documented
in various design studies~\cite{LC}. Not only is the lightest MSSM Higgs
boson observed, but its major decay modes can be measured with high
accuracy, as seen in Fig.~\ref{fig:Battaglia}. Moreover, if sparticles are
light enough to be produced, their masses and other properties can be
measured very precisely, enabling models of supersymmetry breaking to be
tested~\cite{Zerwas}.

As seen in Fig.~\ref{fig:Manhattan}, the sparticles visible at an $e^+e^-$
collider largely complement those visible at the LHC~\cite{benchmark}. In
most of benchmark scenarios proposed, a 1-TeV linear collider would be
able to discover and measure precisely several weakly-interacting
sparticles that are invisible or difficult to detect at the LHC. However,
there are some benchmark scenarios where the linear collider (as well as
the LHC) fails to discover supersymmetry. Only a linear collider with a
higher centre-of-mass energy appears sure to cover all the allowed CMSSM
parameter space, as seen in the lower panels of Fig. 12, which illustrate
the physics reach of a higher-energy lepton collider, such as
CLIC~\cite{CLIC} or a multi-TeV muon collider~\cite{MC}.

\section{Prospects for Other Experiments}

\noindent
\underline{Detection of cold dark matter}\\

\begin{figure}
\epsfxsize180pt 
\figurebox{120pt}{160pt}{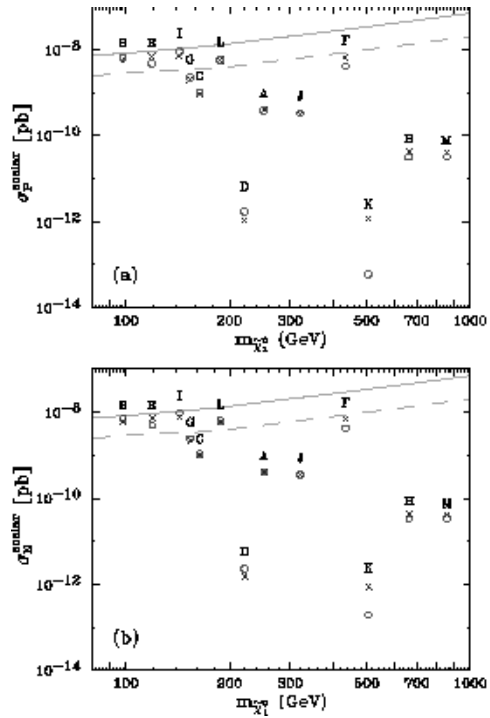}
\caption[]{Rates for the elastic spin-independent scattering  
of supersymmetric relics on (a) protons and (b) neutrons calculated in 
benchmark scenarios~\cite{EFFMO}, compared with upper limits from the 
UKDMC, CDMS and Heidelberg experiments~\cite{KKG}, as well as the range 
suggested by the DAMA collaboration~\cite{DAMA}.}
\label{fig:DM}
\end{figure}

Fig.~\ref{fig:DM} shows rates for the elastic spin-independent scattering
of supersymmetric relics~\cite{EFFMO}, including upper limits from the
UKDMC, CDMS and Heidelberg experiments~\cite{KKG}, as well as the range
suggested by the DAMA collaboration~\cite{DAMA}. Also shown are the rates
calculated in the proposed benchmark scenarios discussed in the previous
section, which are considerably below the DAMA range, but may be within
reach of future projects. Indirect searches for supersymmetric dark matter
via the products of annihilations in the galactic halo or inside the Sun
also have prospects in some of the benchmark scenarios~\cite{EFFMO}.

\noindent
\underline{$\mu\rightarrow e\gamma$ and related processes}\\

\begin{figure}
\epsfxsize190pt
\figurebox{120pt}{160pt}{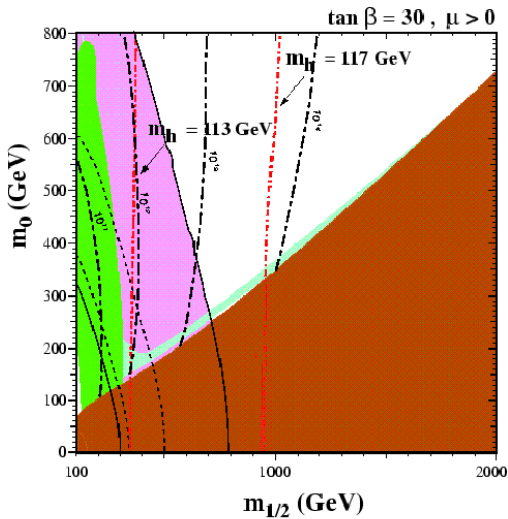}
\caption[]{Illustration in one particular lepton-flavour texture model, 
for $\tan \beta = 30$ and $\mu > 0$,
indicating that $\mu \to e \gamma$ decay may occur at a rate close to 
the present experimental upper limit, in the CMSSM with parameters 
chosen~\cite{CEGL} to match the measured value of 
$g_\mu-2$~\cite{g-2,BNL}.} \label{fig:mueg} \end{figure}

The BNL E821 report of a possible deviation from the Standard Model
suggests that a non-trivial $\mu-\mu-\gamma$ vertex is generated at a
scale $\lappeq$ 1 TeV. Neutrino oscillations indicate that there are
$\Delta L_\mu \not= 0$ processes~\cite{HM}, so it is natural to expect
that there might also be a non-trivial $\mu-e-\gamma$ vertex. This is
indeed the case in a generic supersymmetric GUT, where neutrino mixing
induces slepton mixing~\cite{mueg}. Within this framework, the measurement
of $g_\mu -2$ fixes the sparticle scale, and $\Gamma (\mu\rightarrow
e\gamma)$ may then be calculated within any given flavour texture. Very
approximately, if $g_\mu-2$ is within one or two $\sigma$ of the present
central value, one may expect $B(\mu\rightarrow e\gamma)$ with one or two
orders of magnitude of the present experimental upper limit, as 
illustrated in Fig.~\ref{fig:mueg}~\cite{CEGL}.

The decay $\mu\rightarrow 3e$ and $\mu\rightarrow e$ conversion on nuclei
are expected to occur with branching ratios within two or three orders of
magnitude of $B(\mu\rightarrow e\gamma)$, and it is in principle possible
to measure CP violation in $\mu\rightarrow 3e$ decay. This may provide
another interesting interface with neutrino physics and
cosmology~\cite{CPmu3e}. The minimal supersymmetric seesaw model has six
CP-violating phases: the MNS phase $\delta$, two light-neutrino Majorana
phases, and three phases arising from neutrino Dirac Yukawa couplings,
which may be responsible for our existence via leptogenesis in the early
Universe~\cite{HM}.  The CP-violating neutrino phases induce phases in
slepton mass matrices, which may show up in $\mu\rightarrow 3e$ decay,
$\tau\rightarrow 3e/\mu$ decays and leptonic electric dipole moments. In
principle, the leptogenesis phases might be obtainable by comparing
CP-violating measurements in the charged-lepton and neutrino
sectors~\cite{CPmu3e}.

\noindent
\underline{Proton decay}\\

This could be within reach, with $\tau (p\rightarrow e^+\pi^0)$ via a
dimension-six operator possibly $\sim 10^{35}y$ if $m_{GUT} \sim 10^{16}$
GeV as expected in a minimal supersymmetric GUT. Such a model also
suggests that $\tau (p\rightarrow \bar\nu K^+) < 10^{32} y$ via
dimension-five operators~\cite{dim5}, unless measures are taken to
suppress them~\cite{fsu5}. This provides motivation for a next-generation
megaton experiment that could detect proton decay as well as explore new
horizons in neutrino physics~\cite{UNO}.

\section{Conclusions}

As we have seen, future colliders such as the LHC and a TeV-scale linear
$e^+e^-$ collider have good prospects of discovering supersymmetry and
making detailed measurements. In parallel, $B$ and $\nu$ factories have
good prospects of making inroads on the flavour and unification problems.
Searches for dark matter, stopped-muon experiments and searches for proton
decay also have interesting prospects.

Looking further beyond the Standard Model, how can one hope to test a
Theory of Everything, including quantum gravity?  This should be our
long-term ambition, our analogue of the `faint blue dot' towards which
exoplanetary science is directed, and which motivates much of their
funding. Testing a quantum theory of gravity will be relatively easy if
there are large extra dimensions~\cite{LR,AB}. Much more challenging would 
be
the search for observable effects if the gravitational scale turns out,
after all, to be of the same order as the Planck mass $\sim 10^{19}$ GeV.
Perhaps the only way to reconcile relativity with quantum mechanics is to
modify one or the other, or both~\cite{EMN}? Testing the Theory of
Everything may require thinking beyond the standard `Beyond the Standard
Model' box.

\end{document}